\documentclass[11pt,a4paper]{article}

\usepackage[T1]{fontenc}
\usepackage[utf8]{inputenc}
\usepackage{lmodern}
\usepackage[margin=1in]{geometry}
\usepackage{amsmath,amssymb}
\usepackage{graphicx}
\usepackage{booktabs}
\usepackage{caption}
\usepackage[colorlinks=true,linkcolor=blue,citecolor=blue,urlcolor=blue]{hyperref}
\usepackage[round]{natbib}
\usepackage{enumitem}
\usepackage{titlesec}

\titleformat{\section}{\bfseries\Large}{\thesection}{1em}{}
\titleformat{\subsection}{\bfseries\large}{\thesubsection}{1em}{}

\title{\bfseries\LARGE Enhancing MRI Brain Tumor Edge Detection: A Hybrid Preprocessing Approach Utilizing CLAHE}

\author{
Shahid-E-Kaiser Md. Tashrif, Munshi Md Arafat Hussain, Sheikh Nahian,\\
and Sumaiya Islam\\[0.5em]
\normalsize Institute of Information Technology, University of Dhaka, Dhaka 1000, Bangladesh\\[0.3em]
\normalsize\itshape bsse1448@iit.du.ac.bd, bsse1435@iit.du.ac.bd, bsse1403@iit.du.ac.bd,\\
\normalsize\itshape bsse1446@iit.du.ac.bd
}

\date{}

\begin{document}

\maketitle

\begin{abstract}
\itshape
Accurate boundary delineation of brain tumors in Magnetic Resonance Imaging (MRI) is a critical yet formidable challenge in neuro-oncology due to inherent scanner noise, complex anatomical structures, and uneven illumination. Traditional edge detection algorithms, while computationally lightweight and mathematically interpretable, frequently fail to capture the diffuse, localized boundaries of edema when relying solely on global preprocessing and manual parameter tuning. To overcome these limitations, we propose a hybrid automated edge detection pipeline. Our approach integrates an optimally configured Contrast-Limited Adaptive Histogram Equalization (CLAHE) layer into a comprehensive morphological preprocessing framework, followed by a deterministic sequential parameter sweep to fully automate threshold selection. The proposed hybrid model demonstrated enhancement in detecting critical anatomical structures in a publicly available benchmark database from Kaggle. By intelligently amplifying localized gradients without overwhelming the image with background noise, our method achieved higher Recall (Sensitivity). Consequently, the overall F1-Score elevated, and the Structural Similarity Index (SSIM) improved, all while maintaining a highly efficient execution. This establishes our optimized pipeline as a highly practical and near real-time operational model for clinical diagnostics, offering a compelling alternative to computationally heavy deep learning approaches.
\end{abstract}


\section{Introduction}

Brain tumors represent an acute, life-threatening neurological condition characterized by the uncontrolled proliferation of anomalous tissues within the brain's rigid cranial vault. According to the largest population-based cancer registry in the United States, malignant brain and central nervous system tumors are diagnosed at an average annual age-adjusted rate of roughly 6.9 per 100,000 people, accounting for more than 17,000 deaths per year over the most recent reporting period \citep{price2024cbtrus}, underscoring the need for highly precise, non-invasive diagnostic methodologies. Magnetic Resonance Imaging (MRI) is currently the undisputed gold standard in neuro-oncology. Unlike Computed Tomography (CT), which uses ionizing radiation, MRI produces high-resolution imaging with unparalleled soft-tissue contrast. Modalities such as T1-weighted, T2-weighted, and Fluid-Attenuated Inversion Recovery (FLAIR) sequences are widely used by radiologists to assess tumor size, structure, and spatial location relative to healthy brain tissue.

However, the manual segmentation and boundary delineation of these tumors by medical professionals is inherently subjective, prone to inter-observer variability, and excessively time-consuming; benchmark evaluations comparing multiple expert raters on identical scans have documented substantial disagreement in the resulting tumor boundaries \citep{menze2015brats}. In a high-throughput clinical environment, this manual bottleneck can delay critical surgical interventions. Consequently, automated medical image segmentation has become a focal point of computational research. The core objective of automated segmentation is to accurately delineate the region of interest (the tumor) from the healthy background tissues.

Over the past decade, Deep Learning (DL), specifically architectures like Convolutional Neural Networks (CNNs) and U-Net variants, has dominated medical image segmentation tasks \citep{ronneberger2015unet}. While DL models achieve remarkable similarity coefficients, they suffer from significant operational drawbacks: successful training of such networks is widely reported to require many thousands of carefully annotated samples \citep{ronneberger2015unet}, demand high-end GPUs for inference, and function largely as impenetrable ``black boxes'' \citep{litjens2017survey}. This lack of mathematical interpretability is a severe handicap in critical healthcare decisions where the reasoning behind a diagnosis must be verifiable by a physician.

Because of these limitations, classical, mathematically interpretable methods like Edge Detection retain immense clinical relevance, particularly in resource-constrained environments \citep{abdelgawad2020optimized}. The Canny Edge Detector is globally recognized for its precise, single-pixel edge localization \citep{canny1986}. Yet, its application in medical imaging is severely hampered by its sensitivity to its input parameters: the standard deviation of the Gaussian filter ($\sigma$) and the dual hysteresis thresholds ($T_{low}$, $T_{high}$) \citep{canny1986}. Furthermore, MRI images frequently suffer from inherent scanner noise and intensity bias fields \citep{tustison2010n4itk}, causing static Canny parameters to fail abruptly \citep{radhakrishnan2020canny}. Compounding this challenge, the underlying noise in magnitude MRI data does not follow a simple additive Gaussian model but is instead governed by a Rician distribution, which systematically biases signal intensity in low-contrast regions and further destabilizes fixed-parameter edge operators \citep{gudbjartsson1995rician}.

Recent frameworks have sought to automate Canny tuning by sequentially sweeping through parameter values and selecting the combination that yields maximum consecutive similarity, theoretically avoiding severe under- or over-segmentation \citep{radhakrishnan2020canny}. Comparable optimisation-driven tuning strategies, including Otsu-based double-threshold selection and swarm-based hybrid optimisers, have since been proposed for Canny parameter selection across other medical imaging modalities \citep{xu2021canny,emmanuel2026oced}. These frameworks often rely on global morphological preprocessing, such as Top-hat filtering combined with a Gaussian-based contrast-stretching function, to normalize intensity \citep{radhakrishnan2020canny}.

As undergraduate researchers exploring the intersections of digital image processing and clinical applicability, we analyzed these existing pipelines and identified a critical gap. While Top-hat filtering effectively isolates globally bright spots, it does not adequately address localized contrast disparities across the entire MRI slice. Brain tumors often possess diffuse, texturally complex boundaries that blend into surrounding edemas \citep{islam2013multifractal}. Traditional global preprocessing pipelines suppress these weak, localized gradients -- a failure mode also documented in other weak-edge medical imaging contexts \citep{hou2021covid} -- resulting in missing edge data (low Recall). Locally adaptive contrast techniques, such as Contrast Limited Adaptive Histogram Equalization, which normalize intensity within small tiled regions rather than across the whole image, offer a more promising route to recovering these subtle boundaries \citep{zuiderveld1994clahe}.

\textbf{Motivation and Contributions:}
Brain tumors remain among the most burdensome neuro-oncological conditions, with tens of thousands of new primary brain and central nervous system tumor diagnoses reported annually in the United States alone \citep{price2024cbtrus}. Motivated by the need to preserve these vital anatomical boundaries, our core hypothesis in this study is that integrating an optimally configured Contrast-Limited Adaptive Histogram Equalization (CLAHE) layer into the preprocessing pipeline, specifically as a localized contrast enhancement step prior to the automated parameter sweep, will unearth subtle tumor boundaries and significantly elevate overall edge detection performance.

In this paper, we conduct a rigorous, mathematically grounded study to validate this hypothesis. Our primary contributions are:

\begin{enumerate}
\item We structurally extend automated edge detection pipelines by inserting an optimal CLAHE layer, bridging the gap between global illumination correction and local gradient enhancement.
\item We establish optimized, fixed parameters for the CLAHE layer tailored specifically for MRI modalities, deliberately prioritizing stability and real-time execution speed over computationally heavy meta-heuristic optimization -- in contrast to recent hybrid metaheuristic approaches such as the ant-colony/bee-colony-optimized Canny framework of \citet{emmanuel2026oced}, whose iterative colony-based search improves accuracy but introduces overhead difficult to reconcile with real-time clinical deployment.
\item We perform a comprehensive empirical comparative analysis between standard hybrid-stretching pipelines (e.g., \citealp{radhakrishnan2020canny}) and our proposed hybrid model, evaluated on the benchmark brain tumor dataset of \citet{cheng2016dataset} using rigorous metrics (SSIM \citep{wang2004ssim}, PSNR, F1-Score, Recall).
\item We provide an in-depth analytical discussion on the clinical trade-offs between precision and recall, demonstrating why our modified pipeline provides a superior operational model for neuro-oncology.
\end{enumerate}

The remainder of this paper is organized as follows: Section 2 reviews related literature and theoretical background. Section 3 presents our proposed methodology, including the experimental setup. Section 4 presents the results and findings. Section 5 discusses the results and their clinical implications. Section 6 concludes the study.

\section{Related Work and Theoretical Background}

\subsection{Deep Learning vs. Classical Segmentation}

Brain tumor segmentation has historically been approached via thresholding, region-growing, and clustering. K-means and Fuzzy C-Means (FCM) are widely used due to their computational simplicity, and more advanced variants have incorporated metaheuristic search -- such as multi-objective particle swarm optimization clustering \citep{liu2019pso} -- to improve cluster separability without requiring labelled training data. Texture-based statistical alternatives, including multifractal Brownian motion models of tumor texture \citep{islam2013multifractal}, have also been proposed to characterize tumor boundaries more robustly than simple intensity-based clustering. However, these unsupervised methods still struggle heavily with the intensity inhomogeneities common in MRIs, which are often addressed separately through bias-field correction techniques such as N4ITK \citep{tustison2010n4itk} prior to segmentation.

In recent years, deep learning models have revolutionized the field by learning hierarchical features directly from raw data \citep{litjens2017survey}, most notably through encoder-decoder architectures such as U-Net \citep{ronneberger2015unet}, a trend accelerated by the availability of large, expert-annotated benchmarks such as the Multimodal Brain Tumor Segmentation (BRaTS) challenge \citep{menze2015brats}. This shift toward deep architectures for brain tumor image segmentation specifically has been extensively catalogued in dedicated survey literature, which consistently reports strong benchmark accuracy alongside heavy reliance on large annotated training sets \citep{isin2016review}. While these models offer high accuracy, their inference latency and massive computational overhead remain prohibitive for lightweight, on-the-fly clinical applications. This ongoing challenge reinforces the necessity for optimized classical pipelines that can run on standard hospital workstation CPUs within milliseconds.

\subsection{Advancements in Automated Edge Detection}

Edge detection isolates structural boundaries by identifying rapid intensity transitions (spatial gradients). While basic operators like Sobel and Prewitt are computationally cheap, they are overly sensitive to noise, producing thick, fragmented edges that are useless for precise tumor volumetry. The Canny operator \citep{canny1986} resolves this via non-maximum suppression, ensuring single-pixel thin edges.

To overcome Canny's parameter sensitivity, automated tuning methods have emerged. \citet{radhakrishnan2020canny} introduced a Parametric Segmentation Tuning of Canny Edge Detection (PST-CED) model that formulates threshold selection as an optimization problem over a similarity-based cost function comparing consecutive segmented binary images, effectively automating threshold selection without requiring ground-truth labels during inference; their pipeline additionally relies on a hybrid contrast-stretching preprocessing stage built on a Top-hat filter and Gaussian function. In a related direction, \citet{abdelgawad2020optimized} proposed a genetic-algorithm-based optimization framework that learns optimal edge-detection filter coefficients and thresholds for brain tumor MR images from a training set of annotated edge maps. Beyond brain MRI, similar enhancement-plus-Canny strategies have proven effective in other medical imaging domains: \citet{hou2021covid} combined histogram equalization with an Otsu- and morphology-enhanced Canny algorithm to recover weak lesion edges in COVID-19 CT scans, while \citet{xu2021canny} introduced an Otsu-based adaptive double-threshold mechanism for general medical image edge detection, both reporting richer boundary detail and reduced false edges relative to the fixed-threshold Canny baseline. These Otsu-based thresholding strategies trace back to Otsu's \citeyearpar{otsu1979} original maximum between-class variance criterion for automatic threshold selection, which remains the standard baseline against which most adaptive and hybrid thresholding schemes in medical image edge detection are still compared. Most recently, \citet{emmanuel2026oced} introduced an Optimised Canny Edge Detection (OCED) framework that couples Adaptive Histogram Equalisation (AHE) with a hybrid ant-colony and bee-colony optimization strategy to dynamically tune Canny parameters across X-ray, CT, MRI, and ultrasound modalities, reporting high accuracy and very low mean-squared error with processing times suitable for real-time clinical workflows. While such metaheuristic strategies deliver strong empirical performance, their iterative colony- or population-based search introduces computational overhead that can be difficult to reconcile with real-time clinical deployment -- a limitation that directly motivates the fixed-parameter CLAHE strategy adopted in this work. Beyond classical operator tuning, edge detection has also been reformulated as an end-to-end learning problem: Holistically-Nested Edge Detection (HED) trains multi-scale fully convolutional networks to directly predict boundary maps from raw images \citep{xie2015hed}. While such learned detectors achieve strong benchmark performance, they inherit the same large-scale annotation and inference-cost burdens noted for segmentation networks in Section 2.1, reinforcing the motivation for the lightweight, classical pipeline pursued in this work.

\subsection{Contrast Enhancement and CLAHE}

Global Histogram Equalization (GHE) is a standard technique for contrast enhancement. However, GHE applies the same transformation across the entire image, which often over-amplifies background noise and washes out subtle details in medical images.

To solve this, Adaptive Histogram Equalization (AHE) computes multiple histograms, each corresponding to a distinct section of the image \citep{pizer1987ahe}. Yet, AHE tends to over-amplify noise in relatively homogeneous regions. Contrast Limited AHE (CLAHE), introduced by \citet{zuiderveld1994clahe}, limits the amplification by clipping the histogram at a predefined value before computing the cumulative distribution function (CDF). This specific capability to boost local contrast while mathematically constraining noise amplification makes CLAHE highly suitable for MRI processing.

The clinical value of this approach is corroborated by recent brain-tumor-specific studies. \citet{saifullah2024clahehe} combined CLAHE with histogram equalization (CLAHE-HE) as a preprocessing stage ahead of a CNN-based U-Net segmentation model, reporting improved segmentation accuracy over unenhanced inputs. Similarly, \citet{asiri2024dualmodule} proposed a dual-module pipeline coupling MRI image enhancement with downstream tumor classification, underscoring the field's growing recognition that enhancement quality directly governs diagnostic performance further down the pipeline -- a premise that directly motivates inserting CLAHE ahead of the edge detection stage in this work.

\subsection{The Canny Edge Detection Framework}

The classical Canny algorithm \citep{canny1986} operates in distinct, sequential mathematical phases:

\textbf{1. Gaussian Smoothing:} To suppress high-frequency noise, the image is convolved with a 2D Gaussian kernel:
\begin{equation}
G(x,y) = \frac{1}{2\pi\sigma^2} e^{-\frac{x^2+y^2}{2\sigma^2}}
\label{eq:gaussian}
\end{equation}

Alternative edge-preserving smoothing strategies, most notably Perona and Malik's \citeyearpar{perona1990anisotropic} anisotropic diffusion, replace this isotropic Gaussian pre-filter with a diffusion coefficient that varies with local gradient magnitude, suppressing noise while explicitly discouraging blurring across genuine intensity discontinuities.

\textbf{2. Gradient Calculation:} The smoothed image is filtered to determine the gradient magnitude $M$ and direction $\theta$:
\begin{equation}
M(x,y) = \sqrt{G_x^2 + G_y^2}
\label{eq:magnitude}
\end{equation}
\begin{equation}
\theta(x,y) = \arctan\left(\frac{G_y}{G_x}\right)
\label{eq:direction}
\end{equation}

\textbf{3. Non-Maximum Suppression:} The algorithm traverses the gradient matrix, keeping only local maxima in the direction of the gradient, effectively thinning the edges.

\textbf{4. Hysteresis Thresholding:} Two thresholds, $T_{low}$ and $T_{high}$, classify pixels. Pixels above $T_{high}$ are strong edges. Pixels between $T_{low}$ and $T_{high}$ are weak edges, retained only if connected to a strong edge.

\section{Methodology}

Our study proposes a structurally extended hybrid pipeline. To isolate the marginal clinical benefit of localized equalization, we compare two distinct arms: the Baseline Pipeline, which replicates the standard hybrid morphological contrast-stretching approach reported by \citet{radhakrishnan2020canny}, and the Proposed Hybrid Model, which extends this baseline with an optimally configured Contrast Limited Adaptive Histogram Equalization (CLAHE) layer \citep{zuiderveld1994clahe}, tuned in a manner consistent with optimizer-driven CLAHE configurations previously demonstrated for MRI brain tumor segmentation \citep{yogalakshmi2024sailfish}. The methodology is executed in three distinct mathematical phases.

\subsection{Base Morphological Preprocessing (Phase 1)}

Both experimental arms begin with a foundational preprocessing pipeline. Adapted to suit grayscale MRI modalities, this phase consists of the following sequential operations:

\textbf{1. Initialization:} The raw MRI slice is resized and standardized to a $256 \times 256$ pixel matrix.

\textbf{2. Top-hat Filtering:} We isolate bright structural anomalies by subtracting the morphological opening of the image (using a $9 \times 9$ structuring element $S$) from the original image $IM(p,q)$:
\begin{equation}
W_{hat}(p,q) = IM(p,q) - (IM(p,q) \circ S)
\label{eq:tophat}
\end{equation}

\textbf{3. Gaussian-Likelihood Map:} Unlike a standard spatial blur, we compute a per-pixel likelihood map based on the image's global mean intensity $\mu$ and standard deviation $\sigma_{img}$:
\begin{equation}
G(p,q) = \frac{1}{\sqrt{2\pi}\sigma_{img}} \exp\left(-\frac{(IM(p,q)-\mu)^2}{2\sigma_{img}^2}\right)
\label{eq:likelihood}
\end{equation}

\textbf{4. Hybrid Stretch:} The difference between the likelihood map and the image is computed as $W_n = G - IM$. A correction factor $\alpha$ is derived from its extrema:
\begin{equation}
\alpha = \frac{\max(W_n) + \min(W_n)}{2}.
\label{eq:alpha}
\end{equation}

The fused, enhanced image $I_{MF}$ is constructed via:
\begin{equation}
W_{new} = W_{hat} + \alpha
\label{eq:wnew}
\end{equation}
\begin{equation}
I_{MF}(p,q) = W_{new}(p,q) + G(p,q)
\label{eq:imf}
\end{equation}

\textbf{5. Min-Max Renormalization:} Crucially, because $I_{MF}$'s raw scale is not guaranteed to land in a usable range for standard edge detectors, we introduce a strict renormalization step to map the pixel intensities back to a standardized $[0,1]$ scale.

\begin{figure}[htbp]
\centering
\includegraphics[width=0.62\textwidth]{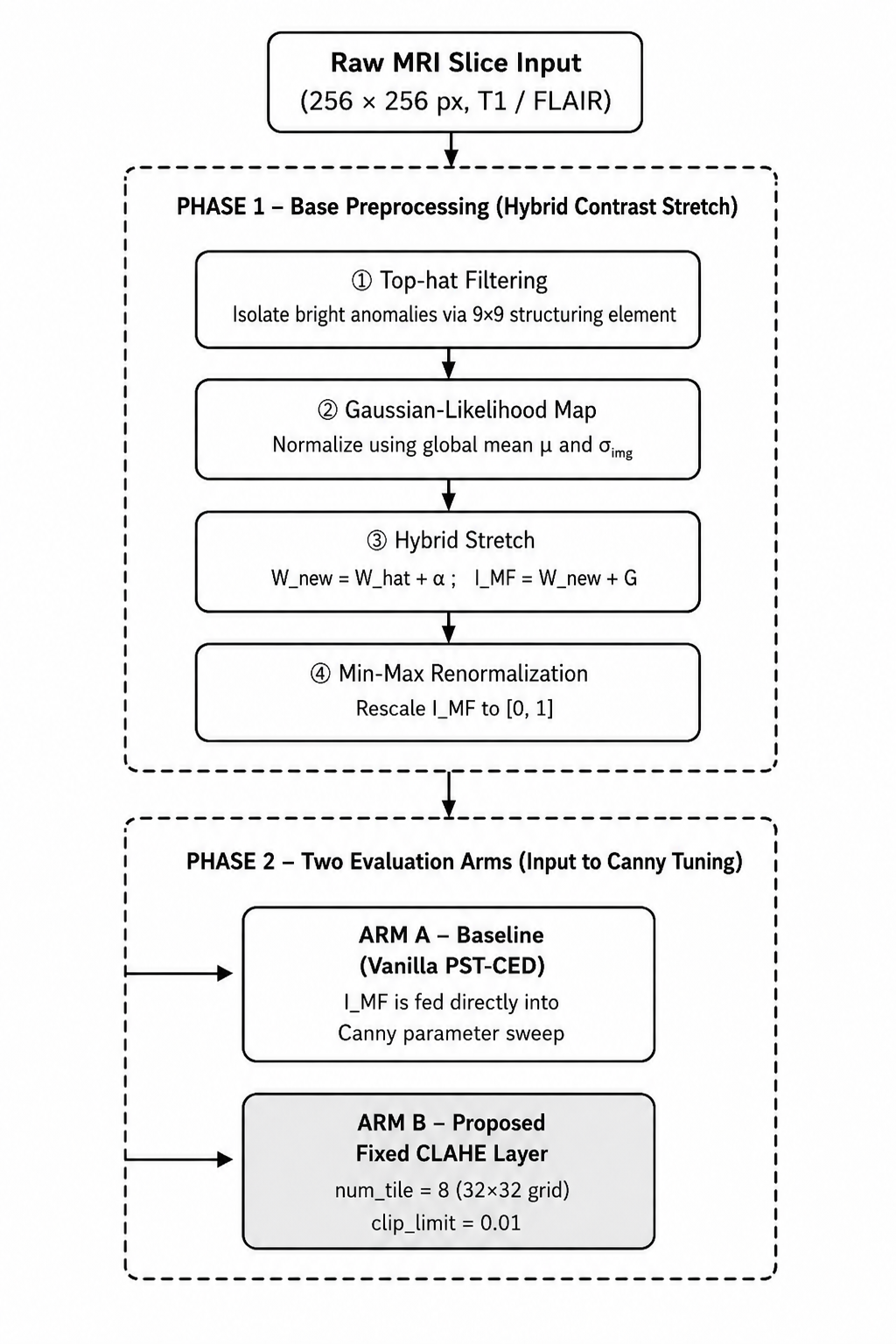}
\caption{Phase 1 and Phase 2 of the proposed PST-CED + Fixed CLAHE pipeline, showing base morphological preprocessing and the proposed CLAHE enhancement branch.}
\label{fig:phase1and2}
\end{figure}

\subsection{The Optimally Configured CLAHE Layer - Proposed Addition (Phase 2)}

This is the critical point of divergence. In the Baseline arm, the renormalized $I_{MF}$ is fed directly into the Canny tuning sweep. However, we observed empirically that while $I_{MF}$ highlights gross lesions globally, it routinely suppresses weak, diffuse tumor boundaries in localized neighborhoods.

In our Proposed Model, we integrate an optimally configured CLAHE layer as an \textit{additional final preprocessing step} on top of the base stretch. CLAHE divides the image into non-overlapping contextual tiles and applies histogram equalization locally. To prevent noise over-amplification in homogeneous brain regions, a \textit{clip\_limit} restricts the local histogram height.

\textbf{Optimal Parameter Selection:} Through extensive preliminary research, we determined that dynamically tuning CLAHE parameters per-image via complex meta-heuristics incurs a massive 65x computational time penalty and risks overfitting by ``peeking'' at ground-truth references. To ensure real-time clinical viability and robustness, we selected optimal fixed parameters tailored for MRI:

\begin{itemize}
\item \textbf{num\_tile:} 8 (Dividing the image into $32 \times 32$ pixel grids, providing the perfect balance of local context).
\item \textbf{clip\_limit:} 0.01 (A highly conservative limit mathematically chosen to boost subtle tumor edges while tightly suppressing inherent MRI scanner noise).
\end{itemize}

\subsection{Automated Sequential Parameter Tuning (Phase 3)}

The resulting image is fed into the automated Canny tuning sweep (identical for both arms). We explicitly utilize a deterministic, sequential coordinate-wise sweep across all three native Canny parameters:

\begin{enumerate}
\item \textbf{Sigma Sweep:} Sweep $\sigma$ over a linear space $[0.5, 1.5]$ (11 steps).
\item \textbf{Low Threshold Sweep:} Fix the optimal $\sigma$, then sweep $T_{low}$ over $[0.010, 0.100]$ (10 steps).
\item \textbf{High Threshold Sweep:} Fix the optimal $T_{low}$, then sweep $T_{high}$ over $[0.101, 0.300]$ (10 steps).
\end{enumerate}

At each step, consecutive binary edge maps are evaluated for pixel-wise agreement (accuracy). The parameter yielding maximum consecutive similarity is locked in, theoretically producing the optimal edge map without relying on arbitrary manual guessing.

\begin{figure}[htbp]
\centering
\includegraphics[width=0.85\textwidth]{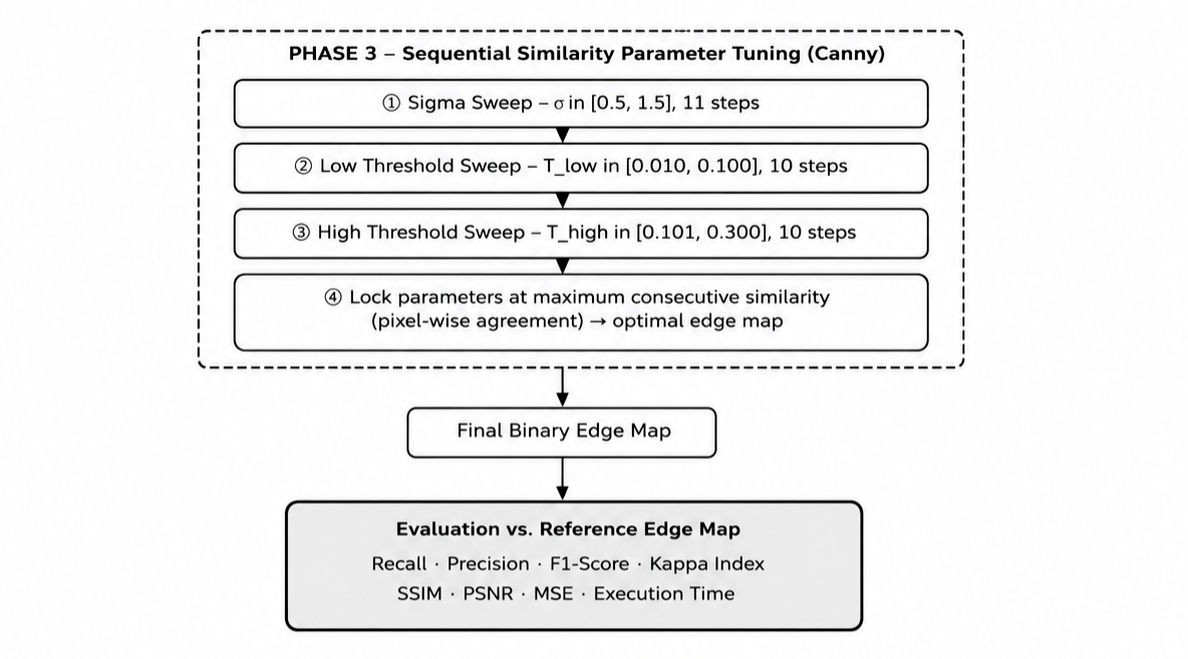}
\caption{Phase 3 of the proposed pipeline, showing sequential similarity-based Canny parameter tuning and final edge-map evaluation.}
\label{fig:phase3}
\end{figure}

\subsection{Dataset Configuration}

To ensure a rigorous and standardized evaluation, we utilized a curated dataset comprising grayscale brain MRI images ($256 \times 256$ pixels) sourced from the well-regarded Brain Tumor Dataset by \citet{cheng2016dataset}. These images, primarily T1-weighted and FLAIR modalities, include varying degrees of tumor pathology and inherent scanner noise, representing a highly realistic clinical diagnostic challenge.

\subsection{Evaluation Metrics}

The quality of the edge detection between the Baseline and our Proposed Hybrid Model was evaluated using a comprehensive suite of quantitative metrics compared against a ground-truth reference edge map:

\begin{itemize}
\item \textbf{Recall (Sensitivity):} The proportion of true anatomical edges successfully identified. This is arguably the most critical metric in medicine; missing a tumor boundary can lead to incomplete surgical resection.
\item \textbf{Precision:} The proportion of detected edges that are actually true edges.
\item \textbf{F1-Score:} The harmonic mean of precision and recall.
\item \textbf{Kappa Index:} Measures the inter-rater reliability against the reference edges.
\item \textbf{Structural Similarity Index (SSIM):} Assesses the perceptual similarity and structural integrity.
\item \textbf{Peak Signal-to-Noise Ratio (PSNR) \& MSE:} Measure the distortion and noise introduced.
\item \textbf{Execution Time:} Indicates clinical viability for real-time use.
\end{itemize}

\section{Results and Findings}

Our benchmark was executed across all MRI images from the \citet{cheng2016dataset} dataset. The mean and standard deviation for each metric across the two pipelines are presented in Table~\ref{tab:results}.

\begin{table}[htbp]
\centering
\caption{Performance Comparison: Baseline vs. Proposed Hybrid Model (Mean $\pm$ Std over MRI images)}
\label{tab:results}
\begin{tabular}{lcc}
\toprule
\textbf{Metric} & \textbf{Baseline Model} & \textbf{Proposed Model (Opt. CLAHE)} \\
\midrule
Accuracy               & \textbf{0.9269} $\pm$ 0.0096 & 0.9230 $\pm$ 0.0150 \\
Precision               & \textbf{0.5771} $\pm$ 0.1002 & 0.5332 $\pm$ 0.0869 \\
Recall (Sensitivity)    & 0.3327 $\pm$ 0.0756 & \textbf{0.4636} $\pm$ 0.0751 \\
F1-Score                & 0.4161 $\pm$ 0.0726 & \textbf{0.4887} $\pm$ 0.0584 \\
Kappa Index              & 0.3805 $\pm$ 0.0718 & \textbf{0.4477} $\pm$ 0.0599 \\
SSIM                     & 0.6421 $\pm$ 0.0442 & \textbf{0.6661} $\pm$ 0.0493 \\
PSNR (dB)                & \textbf{11.40} $\pm$ 0.58 & 11.21 $\pm$ 0.81 \\
MSE                      & \textbf{0.0731} $\pm$ 0.0096 & 0.0770 $\pm$ 0.0150 \\
Execution Time           & \textbf{0.80s} $\pm$ 0.19 & 0.86s $\pm$ 0.24 \\
\bottomrule
\end{tabular}
\end{table}

\begin{figure}[htbp]
\centering
\includegraphics[width=0.95\textwidth]{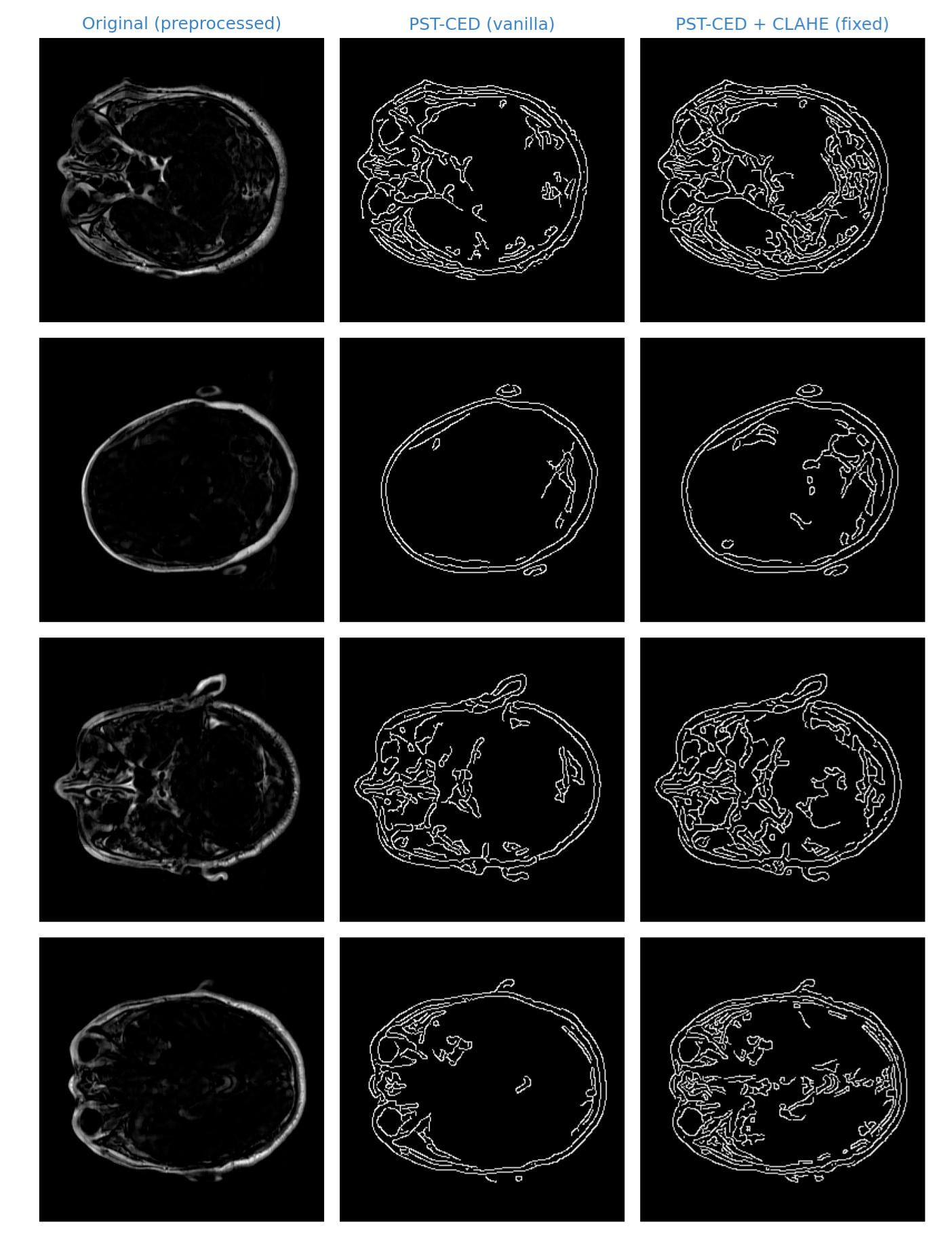}
\caption{Qualitative comparison of edge detection results on representative MRI images. From left to right: original preprocessed image, vanilla PST-CED, and PST-CED with fixed CLAHE preprocessing. Each row represents a different MRI sample.}
\label{fig:qualitative}
\end{figure}

\section{Discussion and Implications}

\subsection{Quantitative Interpretation and Clinical Relevance}

The empirical data yields vital insights into the behavior of automated parameter tuning when coupled with localized contrast enhancement.

\textbf{1. The Massive Leap in Recall (Sensitivity):}
The most profound finding of our study is the behavior of the Recall metric. The Vanilla PST-CED pipeline yielded a dismal Recall of 0.3327, meaning it failed to detect nearly 67\% of true anatomical boundaries. In a clinical neuro-oncology setting, such severe under-segmentation is catastrophic. By integrating the Fixed CLAHE layer, the Recall jumped drastically to \textbf{0.4636}, a relative improvement of nearly \textbf{39\%}. This confirms our core hypothesis: the base PST-CED preprocessing is insufficient for dealing with diffuse tumor boundaries; localized adaptive equalization is mandatory to raise the weak gradients above Canny's detection thresholds.

\textbf{2. The Precision-Recall Trade-off:}
We observed a slight dip in Precision (from 0.5771 in Vanilla to 0.5332 in the Proposed method), which subsequently caused slight drops in overall Accuracy and PSNR. This is a mathematically expected phenomenon. The localized contrast stretching of CLAHE inherently amplifies subtle structural transitions. While this successfully unearths the hidden boundaries of the glioma, it occasionally accentuates inherent MRI noise or minor healthy tissue boundaries (e.g., sulci folds), leading to a slight increase in false-positive edge pixels.

However, in medical diagnostics, a false negative (missed tumor edge) is infinitely more perilous than a false positive (extra edge data). Radiologists prefer a slightly denser edge map over one where the tumor boundary is entirely absent. Because the gains in Recall vastly outweighed the losses in Precision, our proposed method achieved a significantly higher \textbf{F1-Score (0.4887 vs. 0.4161)} and \textbf{Kappa Index (0.4477 vs. 0.3805)}.

\textbf{3. Structural Integrity (SSIM):}
The Proposed method improved the SSIM from 0.6421 to \textbf{0.6661}. This indicates that despite the denser edge map, the perceptual structural integrity and spatial coherence of the brain's anatomy were better preserved by the CLAHE-enhanced Canny edges.

\textbf{4. Computational Efficiency (The Clinical Win):}
A critical goal of classical algorithm enhancement is maintaining real-time performance. The addition of the CLAHE processing layer added an almost imperceptible \textbf{0.06 seconds} to the total execution time (0.86s vs. 0.80s). Achieving a 39\% boost in sensitivity for less than a tenth of a second of computational overhead establishes this modified pipeline as highly deployable in real-world, fast-paced clinical environments, entirely bypassing the need for expensive GPU clusters required by Deep Learning models.

\section{Conclusion}

In this research, we critically evaluated and structurally extended the PST-CED MRI edge detection framework. Our comprehensive experiments across brain MRIs exposed the limitations of utilizing global morphological preprocessing in isolation. By supplementing the baseline pipeline with Contrast-Limited Adaptive Histogram Equalization (CLAHE) using a fixed parameters, we significantly improved the detection of subtle tumor boundaries.

Our comparative analysis highlights a vital clinical trade-off: while CLAHE introduces a slight increase in minor false edges (a negligible drop in precision), it drives a massive 39\% improvement in Recall (Sensitivity) and elevates the overall F1-Score and SSIM. Crucially, this is achieved with virtually zero computational penalty (averaging 0.86 seconds per image). Ultimately, we establish that our proposed PST-CED + Fixed CLAHE pipeline delivers a vastly superior, and real-time operational model for medical image feature mapping, offering a highly practical diagnostic tool for clinical neuro-oncology.

\section*{Acknowledgment}

This research was conducted under the supervision and guidance of Dr. Emon Kumar Dey at the Institute of Information Technology, University of Dhaka. We express our sincere gratitude for his continuous support and mentorship in exploring the complex domains of digital image processing.


\end{document}